\documentclass[5p,authoryear]{elsarticle}

\usepackage{amsmath,amsfonts,amssymb,xcolor}
\usepackage{graphicx}
\usepackage{subfigure}
\usepackage{ulem}
\usepackage{lineno,hyperref}
\hypersetup{colorlinks=true,urlcolor=blue,citecolor=blue}
\modulolinenumbers[5]

\journal{Economics Letters}
\DeclareMathOperator{\argmax}{argmax}

\def\half{{\textstyle\frac{1}{2}}}

\def\Rset{\mathbb{R}}
\DeclareMathOperator{\rk}{rk}

\newcommand{\bm}[1]{\mbox{\boldmath $#1$}} 
\newcommand{\what}[1]{\widehat{#1}}
\newcommand*{\defeq}{\mathrel{\vcenter{\baselineskip0.5ex \lineskiplimit0pt
                     \hbox{\scriptsize.}\hbox{\scriptsize.}}}%
                     =}

\newtheorem{thm}{Theorem}

\newtheorem{proposition}[thm]{Proposition}
\newdefinition{rmk}{Remark}
\newproof{pf}{Proof}

\begin{document}

\begin{frontmatter}

\title{Bilinear form test statistics for extremum estimation}
\tnotetext[titlenote]{This research was supported by Fondecyt grant 11140433,
Regione Autonoma della Sardegna Master and Back grant PRR-MAB-A2011-24192
(F. Crudu) and Fondecyt grant 1140580 (F. Osorio).}

\author[DEPS]{Federico Crudu\corref{cor1}%
  \fnref{fn1}}
\ead{federico.crudu@unisi.it}

\author[DMAT]{Felipe Osorio\fnref{fn2}}
\ead{felipe.osorios@usm.cl}

\cortext[cor1]{Correspondence to: Department of Economics and Statistics, University
  of Siena, Piazza San Francesco, 7/8 53100 Siena, Italy.}
\address[DEPS]{Department of Economics and Statistics, University of Siena, Italy}
\address[DMAT]{Departamento de Matem\'{a}tica, Universidad T\'{e}cnica Federico Santa Mar\'{i}a, Chile}

\begin{abstract}
  This paper develops a set of test statistics based on bilinear forms in the context of
  the extremum estimation framework with particular interest in nonlinear hypothesis.
  We show that the proposed statistic converges to a conventional chi-square limit.
  A Monte Carlo experiment suggests that the test statistic works well in finite samples.
\end{abstract}


\end{frontmatter}


\section{Introduction}\label{sec:Intro}

The purpose of this paper is to introduce a novel test statistic for extremum estimation
(EE). In this very general setting \citep[see for instance][]{Gourieroux:1995, Hayashi:2000},
conventional test statistics are defined either in terms of differences (pseudo likelihood
ratio or distance statistic) or in terms of quadratic forms (Wald, Lagrange multiplier
also known as Rao's \citeyearpar{Rao:1948} score statistic). The test proposed in this
paper is defined in terms of a bilinear form ($BF$). This approach is not entirely new
as a bilinear form test for maximum likelihood was introduced by \cite{Terrell:2002}
\citep[see also the monograph by][]{Lemonte:2016}. Our test statistic has a conventional
chi-square limit and, similarly to the Wald test, it is generally not invariant to the
definition of the null hypothesis. It is, though, easy to see that in the context of
linear models the $BF$ test is equal to the distance statistic, which is, on the other
hand, invariant. Furthermore, when nonlinear models are involved our Monte Carlo simulations
suggest that the discrepancy induced by equivalent definitions of the null hypothesis
is relatively small when compared, e.g., to the Wald test. In the general case, the
computational burden associated to the $BF$ statistic is comparable to that of the
distance metric statistic, since both the estimator under the null and under the alternative
must be calculated. To the best of our knowledge this is the first paper that deals with
this problem in the context of EE.

The remainder of the paper unfolds as follows. Section \ref{sec:tests} contains the
description of the test statistics for a generic, potentially nonlinear, null hypothesis
and their asymptotic properties; the asymptotic results and the corresponding proofs are
presented in a concise fashion and are mostly based on the results in \cite{Gourieroux:1995}.
In Section \ref{sec:experiments} we study, via Monte Carlo experiments, the finite
sample properties of the test in comparison with other more conventional EE test
statistics. Section \ref{sec:conclusion} offers some conclusions while the appendices
contain the proofs of the asymptotic results.

\section{A bilinear form test statistic}\label{sec:tests}

Let us consider a scalar objective function $Q_n(\bm{\beta})$ that depends on a set
of data ${\bm{w}_i}, i=1,\dots,n$ with ${\bm{w}_i}\in\Rset^k$ and $\bm{\beta}\in\mathcal{B}
\subset\Rset^p$ where $\mathcal{B}$ is compact. The EE for our objective function
can be defined as
\begin{equation}\label{eq:UN-EE}
  \what{\bm{\beta}}_n = \underset{\beta\in\mathcal{B}}{\argmax}\, Q_n(\bm{\beta}).
\end{equation}
Let us now suppose that we want to test the following null hypothesis
\begin{equation}\label{eq:null}
  H_0: \bm{g}(\bm{\beta}_0) = \bm{0}
\end{equation}
given that $\bm{g}:\Rset^p\to\Rset^q$ is a continuously differentiable function and
$\bm{G}(\bm{\beta}) = \partial\bm{g}(\bm{\beta})/\partial\bm{\beta}^\top$ is a $q\times p$
matrix with $\rk(\bm{G}(\bm{\beta})) = q$. The resulting constrained estimator is
defined as the solution of the Lagrangian problem
\begin{equation}\label{eq:objective}
  L_n(\bm{\beta},\bm{\lambda}) = Q_n(\bm{\beta}) - \bm{g}^\top(\bm{\beta})\bm{\lambda},
\end{equation}
where $\bm{\lambda}$ denotes a vector of Lagrange multipliers. Hence,
\begin{equation}\label{eq:constr-EE}
  \widetilde{\bm{\beta}}_n = \underset{\beta\in\{\mathcal{B}: g(\beta) = 0\}}{\argmax}
  L_n(\bm{\beta},\bm{\lambda}).
\end{equation}
The null hypothesis in Equation (\ref{eq:null}) can be tested, for example, by means
of the simple Wald ($W$) test, that only requires the unconstrained estimator or either
the Lagrange multiplier ($LM$) test or the distance metric ($D$) statistic that both
require the constrained estimator in Equation (\ref{eq:constr-EE}). The $BF$ tests
that we propose are generalizations of Terrell's gradient statistic \citep{Terrell:2002}
to the EE context.\footnote{Sometimes the term \emph{gradient statistic} is used to
indicate the $LM$ test for GMM \citep[see for example Chapter 22 in][]{Ruud:2000}.
To avoid confusion we prefer the expression \emph{bilinear form test} and the corresponding
abbreviation $BF$.} Let us first define $\bm{A}_n(\bm{\beta}_0)\defeq \partial^2
Q_n(\bm{\beta}_0)/\partial\bm{\beta}\partial\bm{\beta}^\top$ and assume that $\bm{A}_n(\bm{\beta}_0)
\stackrel{\sf a.s.}{\to} \bm{A}$ uniformly. Let us also assume that
\[
 \sqrt{n}\,\frac{\partial
  Q_n(\bm{\beta}_0)}{\partial\bm{\beta}}\stackrel{\sf D}{\to} \mathsf{N}_p(\bm{0},
  \bm{B}).
\]
Furthermore, let $\bm{G}\defeq\bm{G}(\bm{\beta}_0)$, $\bm{S} = \bm{G}\{-\bm{A}\}^{-1}
\bm{G}^\top$ and $\bm{\Omega} = \bm{G}\bm{A}^{-1}\bm{B}\bm{A}^{-1}\bm{G}^\top$. Then,
\begin{equation}\label{eq:BFstat1}
  BF_1 \defeq n\widetilde{\bm{\lambda}}{}_n^\top\bm{S}\bm{\Omega}^{-1}\bm{g}(\what{\bm{\beta}}_n)
\end{equation}
where $\widetilde{\bm{\lambda}}{}_n$ is the solution for $\bm{\lambda}$ in the Lagrangian
problem defined by Equation (\ref{eq:objective}). The $BF$ statistic also has the following
alternative formulations. Let $\bm{G}^+= \bm{G}^\top\{\bm{G}\bm{G}^\top\}^{-1}$ denote the
Moore-Penrose inverse of $\bm{G}$ \citep[see, for instance,][p.\,38]{Magnus:2007}. Then,
\begin{align}
  BF_2 & \defeq n\frac{\partial Q_n(\widetilde{\bm{\beta}}_n)}{\partial\bm{\beta}^\top}
  \bm{G}^+\bm{S}\bm{\Omega}^{-1}\bm{g}(\what{\bm{\beta}}_n) \label{eq:BFstat2} \\
  BF_3 & \defeq n\frac{\partial Q_n(\widetilde{\bm{\beta}}_n)}{\partial\bm{\beta}^\top}
  \bm{G}^+\bm{S}\bm{\Omega}^{-1}\bm{G}(\what{\bm{\beta}}_n-\widetilde{\bm{\beta}}_n). \label{eq:BFstat3}
\end{align}
Let us define $\bm{P}_G \defeq \bm{G}^+\bm{G}$ and assume that $\bm{B} = -\bm{A}$, which
leads to $\bm{S} = \bm{\Omega}$. We then obtain the following specifications:
\begin{align}
  BF_4 & \defeq n\widetilde{\bm{\lambda}}{}_n^\top\bm{g}(\what{\bm{\beta}}_n)\label{eq:BFstat4} \\
  BF_5 & \defeq n\frac{\partial Q_n(\widetilde{\bm{\beta}}_n)}{\partial\bm{\beta}^\top}
  \bm{G}^+\bm{g}(\what{\bm{\beta}}_n) \label{eq:BFstat5} \\
  BF_6 & \defeq n\frac{\partial Q_n(\widetilde{\bm{\beta}}_n)}{\partial\bm{\beta}^\top}
  \bm{P}_G(\what{\bm{\beta}}_n-\widetilde{\bm{\beta}}_n) \label{eq:BFstat6} \\
  BF_7 & \defeq n\frac{\partial Q_n(\widetilde{\bm{\beta}}_n)}{\partial\bm{\beta}^\top}
  (\what{\bm{\beta}}_n-\widetilde{\bm{\beta}}_n). \label{eq:BFstat7}
\end{align}
The assumption that $\bm{B} = -\bm{A}$ is not very restrictive as it may include as
special cases maximum likelihood and GMM statistics \citep[see][Chapter\,7]{Hayashi:2000}.
Next, we consider a quadratic objective function where this condition is satisfied.
\begin{rmk}
  Let $Q_n(\bm{\beta}) = -\half\bm{f}_n^\top(\bm{\beta})\bm{W}^{-1}\bm{f}_n(\bm{\beta})$ where
  $\bm{f}_n(\bm{\beta})$ is a set of sample moment conditions and $\bm{W}$ is a conformable
  positive definite matrix, then
  \[
    \frac{\partial Q_n(\bm{\beta})}{\partial\bm{\beta}} = -\bm{F}_n^\top(\bm{\beta})\bm{W}^{-1}
    \bm{f}_n(\bm{\beta}),
  \]
  with $\bm{F}_n(\bm{\beta}) = \partial\bm{f}_n(\bm{\beta})/\partial\bm{\beta}^\top$, and
  \[
    \frac{\partial^2 Q_n(\bm{\beta})}{\partial\bm{\beta}\partial\bm{\beta}^\top} =
    -\Big[\frac{\partial\bm{F}_n^\top(\bm{\beta})}{\partial\bm{\beta}}\Big]
    \big[\bm{W}^{-1}\bm{f}_n(\bm{\beta})\big] - \bm{F}_n^\top(\bm{\beta})\bm{W}^{-1}\bm{F}_n(\bm{\beta}),
  \]
  where $[\cdot][\cdot]$ denotes array multiplication \citep[See Appendix A.2 of]
  [for details]{Wei:1998}. If $\bm{f}_n(\bm{\beta}_0)$ converges to its expected
  value, i.e. zero, its derivatives converge almost surely to finite full rank matrices and $\sqrt{n}\bm{f}_n(\bm{\beta}_0)
  \stackrel{\sf D}{\to} \mathsf{N}(\bm{0},\bm{W})$, then we find $\bm{B} = \bm{F}^\top
  \bm{W}^{-1}\bm{F}$ and $\bm{A} = -\bm{F}^\top\bm{W}^{-1}\bm{F}$. Hence, $\bm{B}=-\bm{A}$ holds.
\end{rmk}

The following proposition shows that the $BF$ tests are asymptotically equivalent
and have a conventional chi-square limit.

\begin{proposition}\label{result:main}
  Under the assumptions of Property 24.16 and Property 24.10 in \cite{Gourieroux:1995}, with
  $\bm{g}:\Rset^p\to\Rset^q$ being a continuously differentiable function and $\bm{G}
  (\bm{\beta}) = \partial\bm{g}(\bm{\beta})/\partial\bm{\beta}^\top$ a $q\times p$
  matrix with $\rk(\bm{G}(\bm{\beta})) = q$,
  \[
    BF_k\stackrel{\sf D}{\to}\chi^2_q, \qquad k=1,2,3.
  \]
  If, in addition, $\bm{B} = -\bm{A}$ holds, then
  \[
    BF_k\stackrel{\sf D}{\to}\chi^2_q, \qquad k=4,5,6,7.
  \]
\end{proposition}
\begin{pf}
  See \ref{app:proof}.
\end{pf}

\begin{rmk}
  When $Q_n(\bm{\beta}) = \overline{\ell}_n(\bm{\beta})$ is the log-likelihood
  function we obtain that the $BF$ statistic is given by
  \begin{equation}\label{eq:BF-logLik}
    BF = \bm{U}_n^\top(\widetilde{\bm{\beta}}_n)\bm{G}^+\bm{g}(\what{\bm{\beta}}_n),
  \end{equation}
  where $\bm{U}_n(\bm{\beta}) = \partial\overline{\ell}_n(\bm{\beta})/\partial\bm{\beta}$
  denotes the score function. We must highlight that (\ref{eq:BF-logLik}) is an extension
  of the test proposed by \cite{Terrell:2002} to tackle nonlinear hypotheses.
\end{rmk}

\begin{rmk}
  It is interesting to see that in the case of the linear model, $D$ and $BF$ are equal.
  Let us consider, the example in \cite{Hansen:2006}. The $BF$ statistic is
  \[
    BF = (\bm{y} - \bm{X}\widetilde{\bm{\beta}}_n)^\top\bm{X}\bm{B}^{-1}\bm{X}^\top
    \bm{X}(\what{\bm{\beta}}_n - \widetilde{\bm{\beta}}_n).
  \]
  Since $\what{\bm{\beta}}_n = (\bm{X}^\top\bm{X})^{-1}\bm{X}^\top\bm{y}$ and $\bm{X}^\top
  (\bm{y} - \bm{X}\widehat{\bm{\beta}}_n) = \bm{0}$, it follows immediately that $BF = D$.
\end{rmk}

Next proposition establishes the asymptotic equivalence between $BF$ and $LM$ tests
for nonlinear hypothesis \citep[a discussion about $LM$ statistics under general
settings can be found in][]{Boos:1992}.

\begin{proposition}\label{result:equiv}
  The $BF$ test statistic in Equation (\ref{eq:BFstat1}) and the Lagrange multiplier
  test statistic
  \[
    LM \defeq n\widetilde{\bm{\lambda}}{}_n^\top\bm{S}\bm{\Omega}^{-1}\bm{S}\widetilde{\bm{\lambda}}_n,
  \]
  are asymptotically equivalent under $H_0:\bm{g}(\bm{\beta}_0) = \bm{0}$. Their
  common asymptotic distribution is $\chi^2_q$.
\end{proposition}

\begin{pf}
  See \ref{app:equiv}.
\end{pf}

\begin{table*}[!htp]
  \begin{center}
  \caption{Empirical size for a 5\% test. The superscripts $A$ and $B$ refer to the
  fact that $W$ and $BF$ are computed using the null hypotheses in Equations (\ref{eq:H0A})
  and (\ref{eq:H0B}), respectively.}\label{tab:exp1}
  \begin{tabular}{llrcccccc} \hline\hline
    Scenario & $(\beta_2,\beta_3)$ & $n$ & $W^A$ & $W^B$ & $BF^A$ & $BF^B$ & $LM$ & $D$ \\ \hline
    I        & (10,0.1)  &  20 & 0.420 & 0.176 & 0.067 & 0.064 & 0.084 & 0.087 \\
             &           &  50 & 0.282 & 0.106 & 0.065 & 0.059 & 0.068 & 0.074 \\
             &           & 100 & 0.197 & 0.077 & 0.059 & 0.059 & 0.061 & 0.061 \\
             &           & 500 & 0.104 & 0.052 & 0.048 & 0.048 & 0.049 & 0.049 \\ \hline
    II       & (5,0.2)   &  20 & 0.277 & 0.178 & 0.068 & 0.065 & 0.083 & 0.086 \\
             &           &  50 & 0.171 & 0.108 & 0.058 & 0.057 & 0.067 & 0.068 \\
             &           & 100 & 0.127 & 0.078 & 0.058 & 0.058 & 0.061 & 0.062 \\
             &           & 500 & 0.070 & 0.052 & 0.047 & 0.047 & 0.048 & 0.048 \\ \hline
    III      & (2,0.5)   &  20 & 0.145 & 0.175 & 0.066 & 0.067 & 0.082 & 0.082 \\
             &           &  50 & 0.096 & 0.113 & 0.062 & 0.057 & 0.070 & 0.075 \\
             &           & 100 & 0.078 & 0.082 & 0.056 & 0.056 & 0.062 & 0.062 \\
             &           & 500 & 0.049 & 0.055 & 0.045 & 0.045 & 0.050 & 0.050 \\ \hline
    IV       & (1,1)     &  20 & 0.140 & 0.170 & 0.084 & 0.070 & 0.086 & 0.101 \\
             &           &  50 & 0.095 & 0.108 & 0.062 & 0.062 & 0.070 & 0.070 \\
             &           & 100 & 0.074 & 0.080 & 0.066 & 0.066 & 0.067 & 0.067 \\
             &           & 500 & 0.055 & 0.055 & 0.056 & 0.056 & 0.061 & 0.061 \\ \hline\hline
  \end{tabular}
  \end{center}
\end{table*}

\section{Monte Carlo simulations}\label{sec:experiments}

To study the finite sample properties of the $BF$ statistic we consider two equivalent
nonlinear null hypotheses, as in \cite{Gregory:1985} \citep[see also][]{Hansen:2006,
Lafontaine:1986}. The $BF$ test, which is not invariant to the specification of the
null, is compared against the $W$, $LM$ and $D$ statistics. While the first test is
known to be not invariant, the last two tests are invariant and work well in finite
samples (see, for instance, \citealt{Dagenais:1991} and \citealt{Hansen:2006}). The
performance of the tests is measured in terms of how close the empirical size is to
the 5\% nominal size and in terms of the discrepancy between the empirical sizes produced
by competing equivalent hypotheses. Here, the distance metric statistic $D$ is defined as
\[
  D \defeq n(Q_n(\widetilde{\bm{\beta}}_n) - Q_n(\what{\bm{\beta}}_n)),
\]
where $Q_n(\bm{\beta})$ is the objective function of the nonlinear least squares
estimator. In our experiment the $BF$ statistic defined in Equation (\ref{eq:BFstat7})
was used.
\begin{figure*}[!ht]
  \centering
  \subfigure[$n = 20$]{
    \includegraphics[width = 0.26\linewidth]{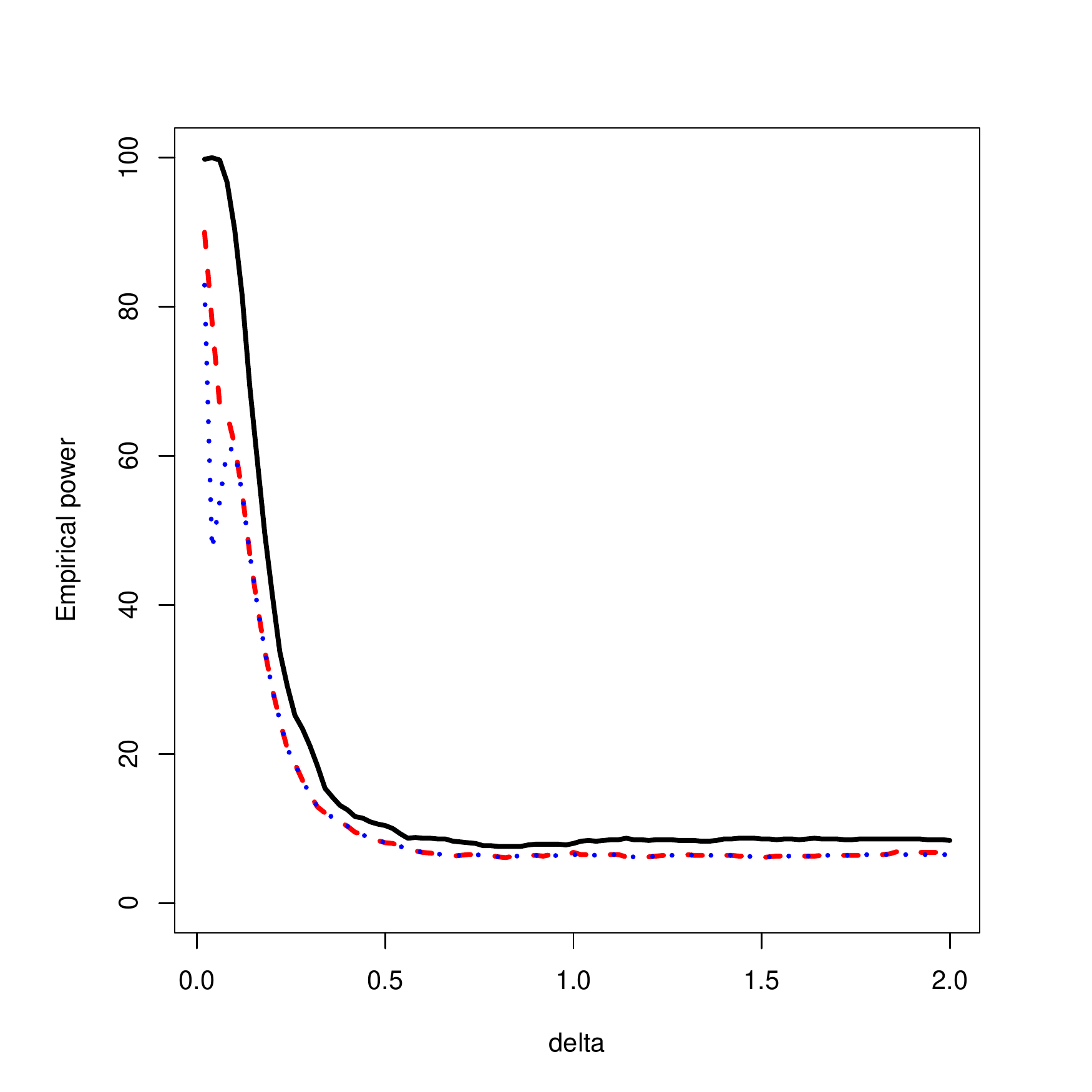}
  }
  \subfigure[$n = 50$]{
    \includegraphics[width = 0.26\linewidth]{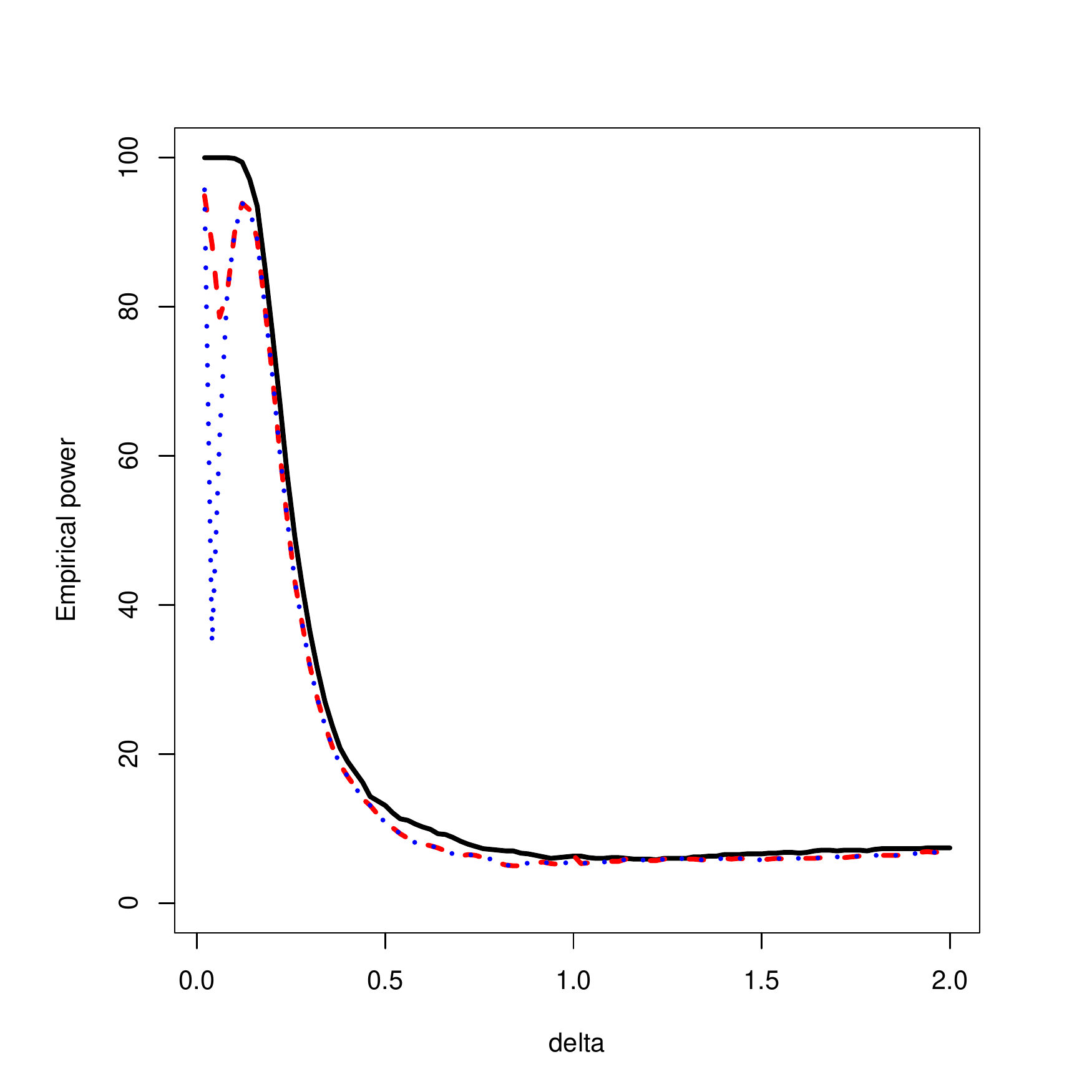}
  }
  \subfigure[$n = 100$]{
    \includegraphics[width = 0.26\linewidth]{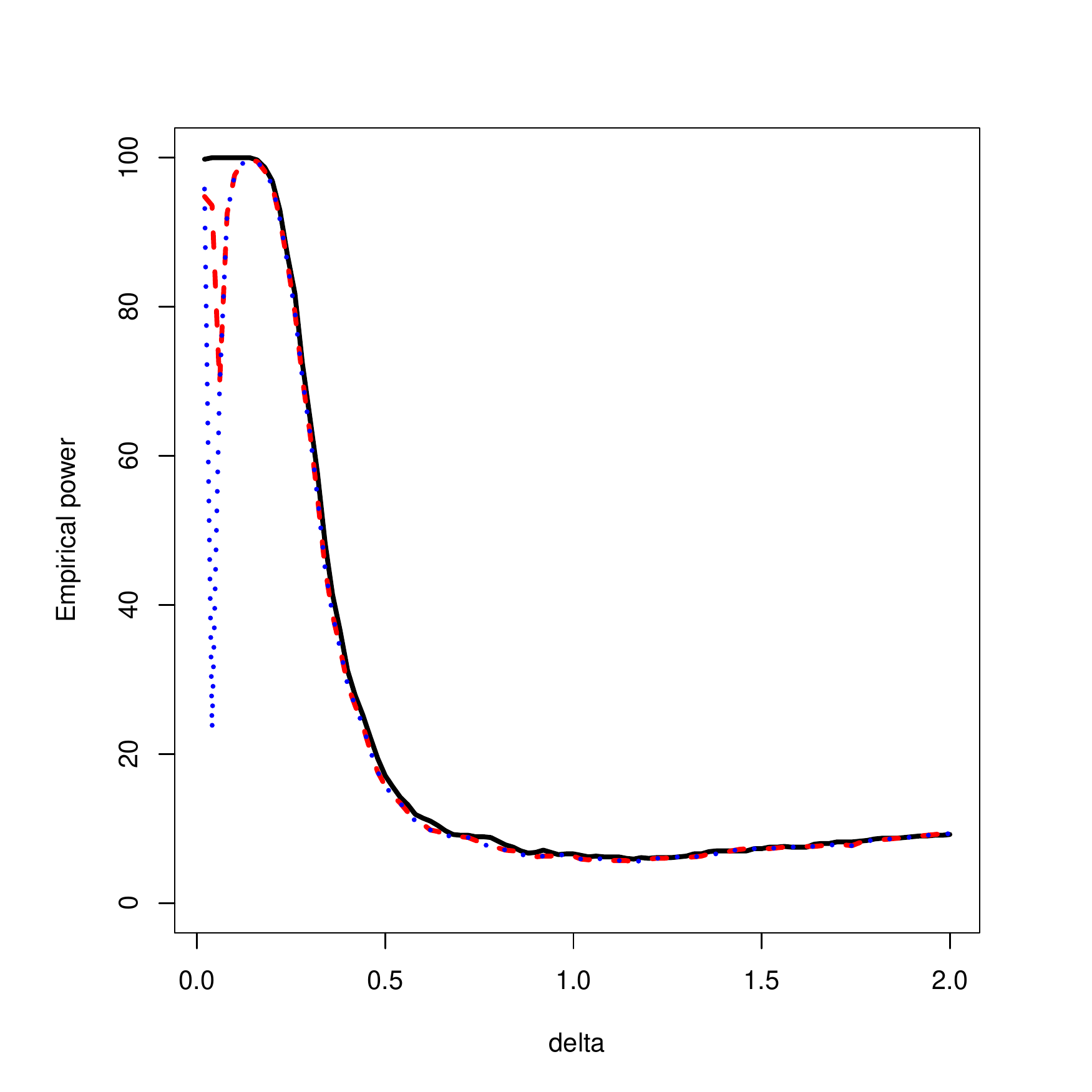}
  }
  \subfigure[$n = 20$]{
    \includegraphics[width = 0.26\linewidth]{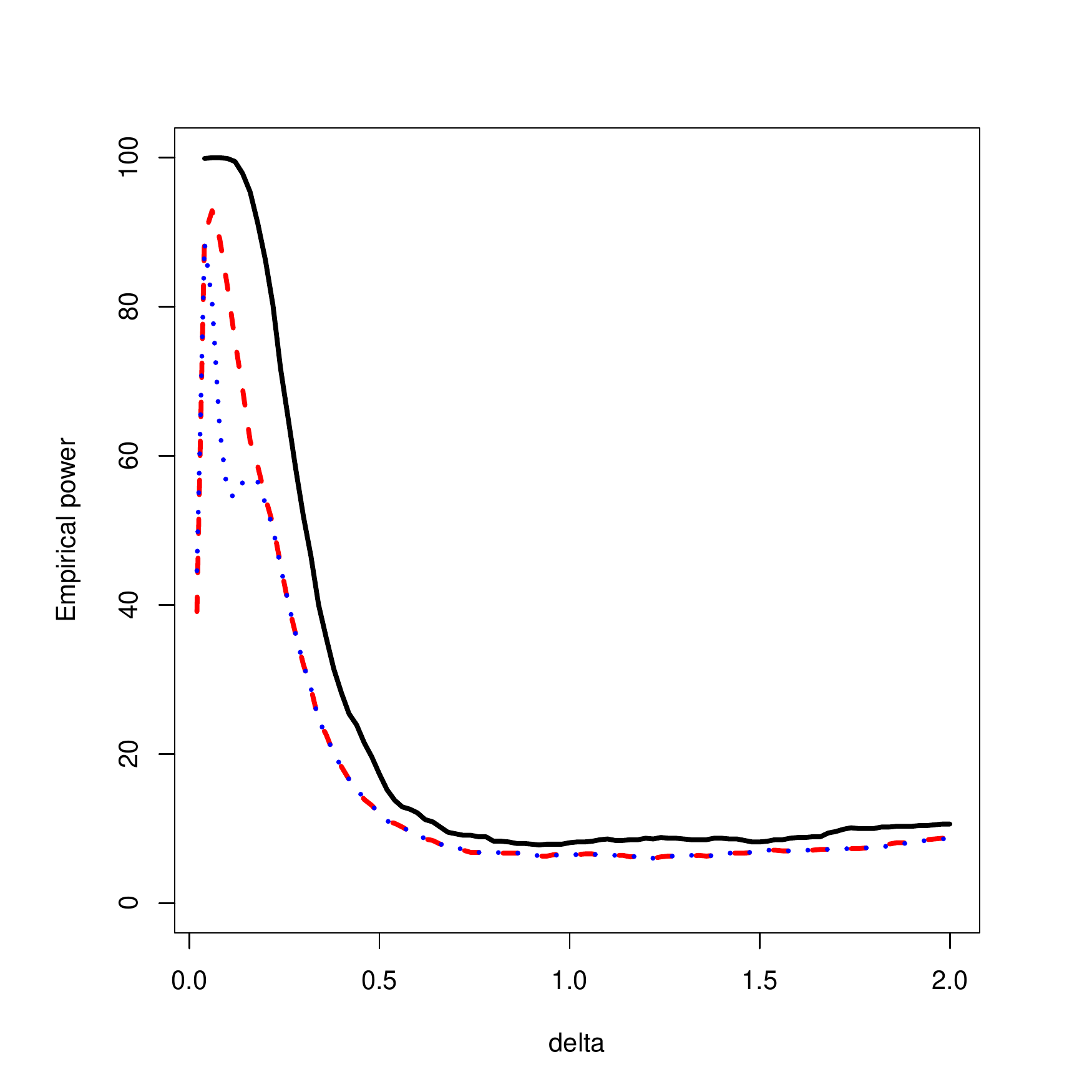}
  }
  \subfigure[$n = 50$]{
    \includegraphics[width = 0.26\linewidth]{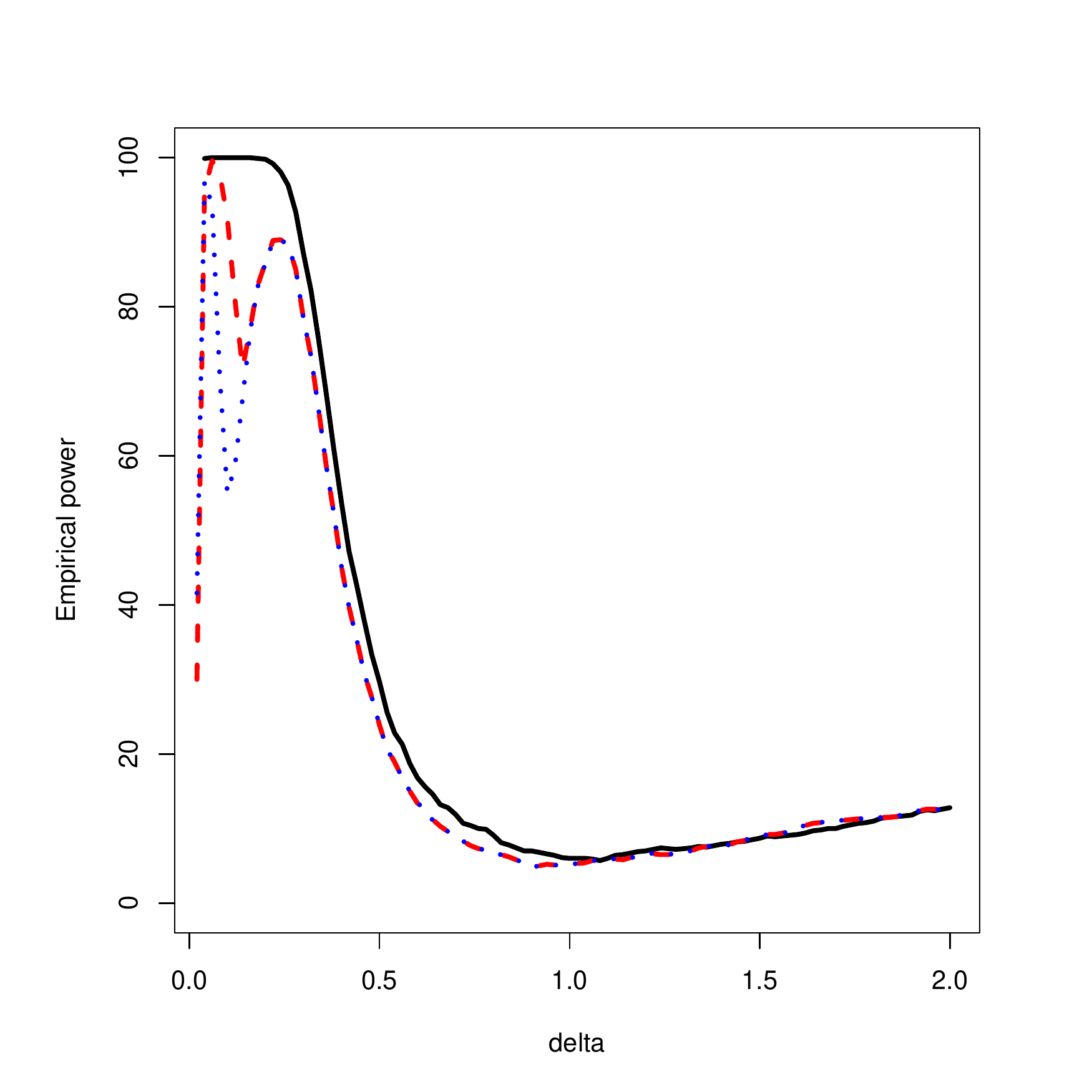}
  }
  \subfigure[$n = 100$]{
    \includegraphics[width = 0.26\linewidth]{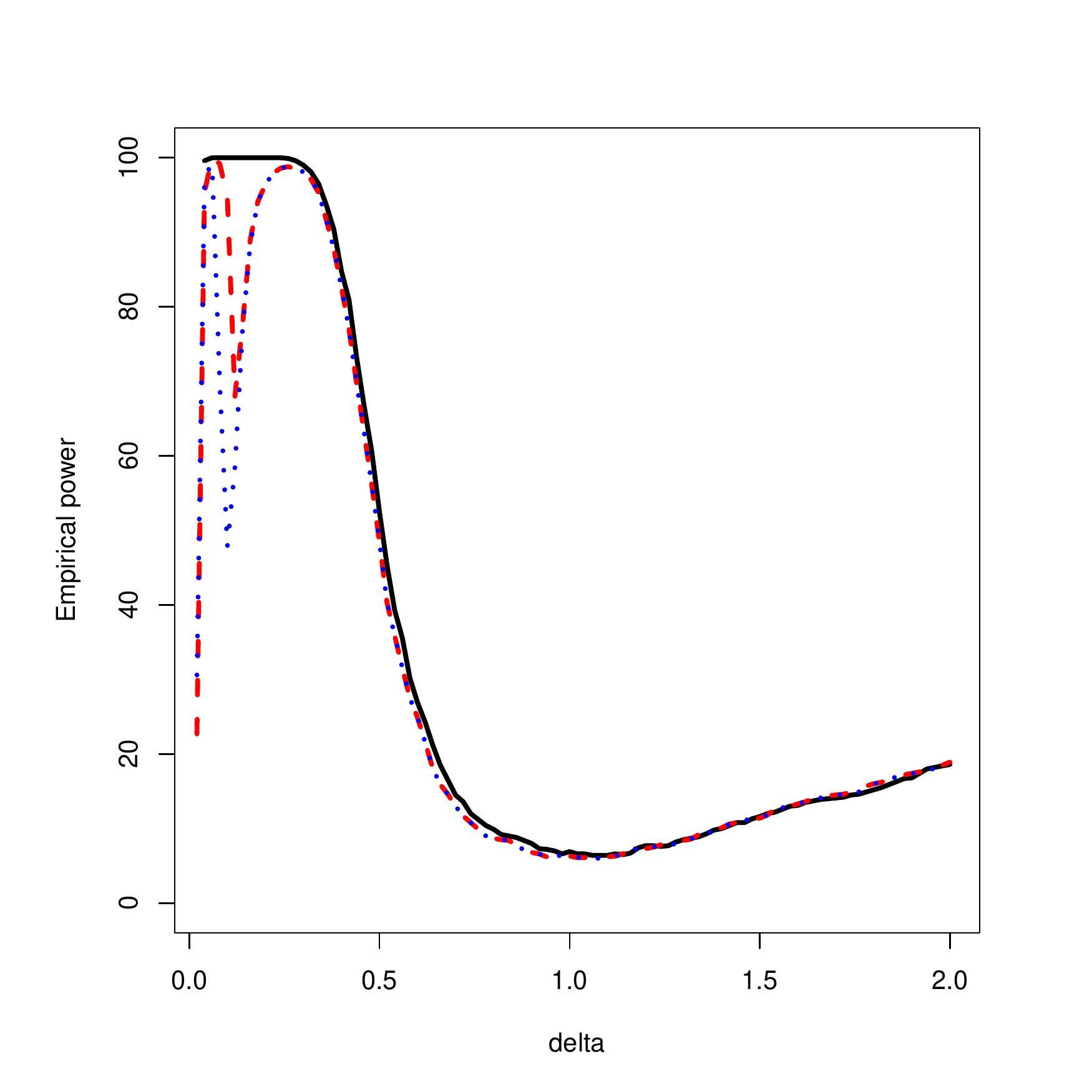}
  }
  \subfigure[$n = 20$]{
    \includegraphics[width = 0.26\linewidth]{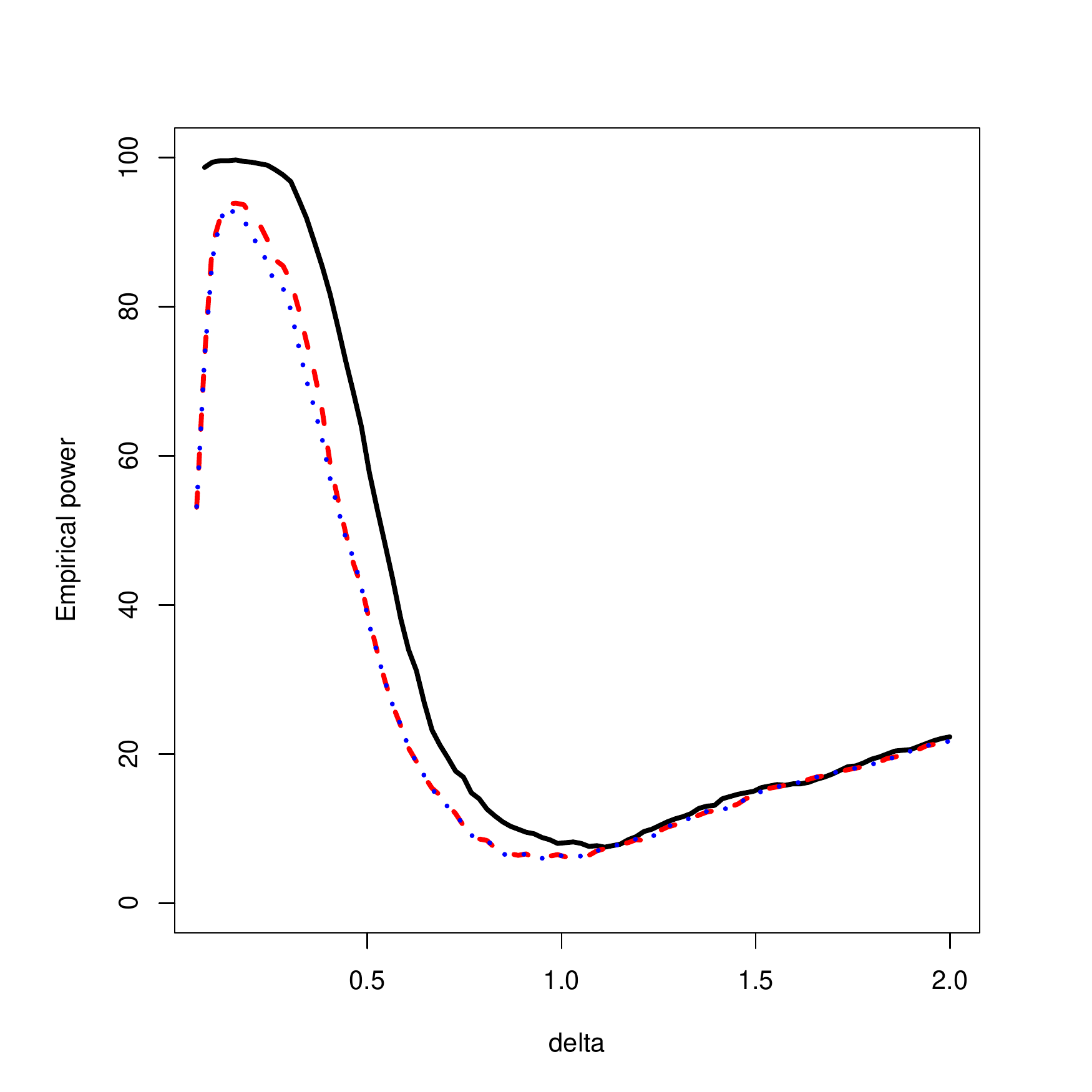}
  }
  \subfigure[$n = 50$]{
    \includegraphics[width = 0.26\linewidth]{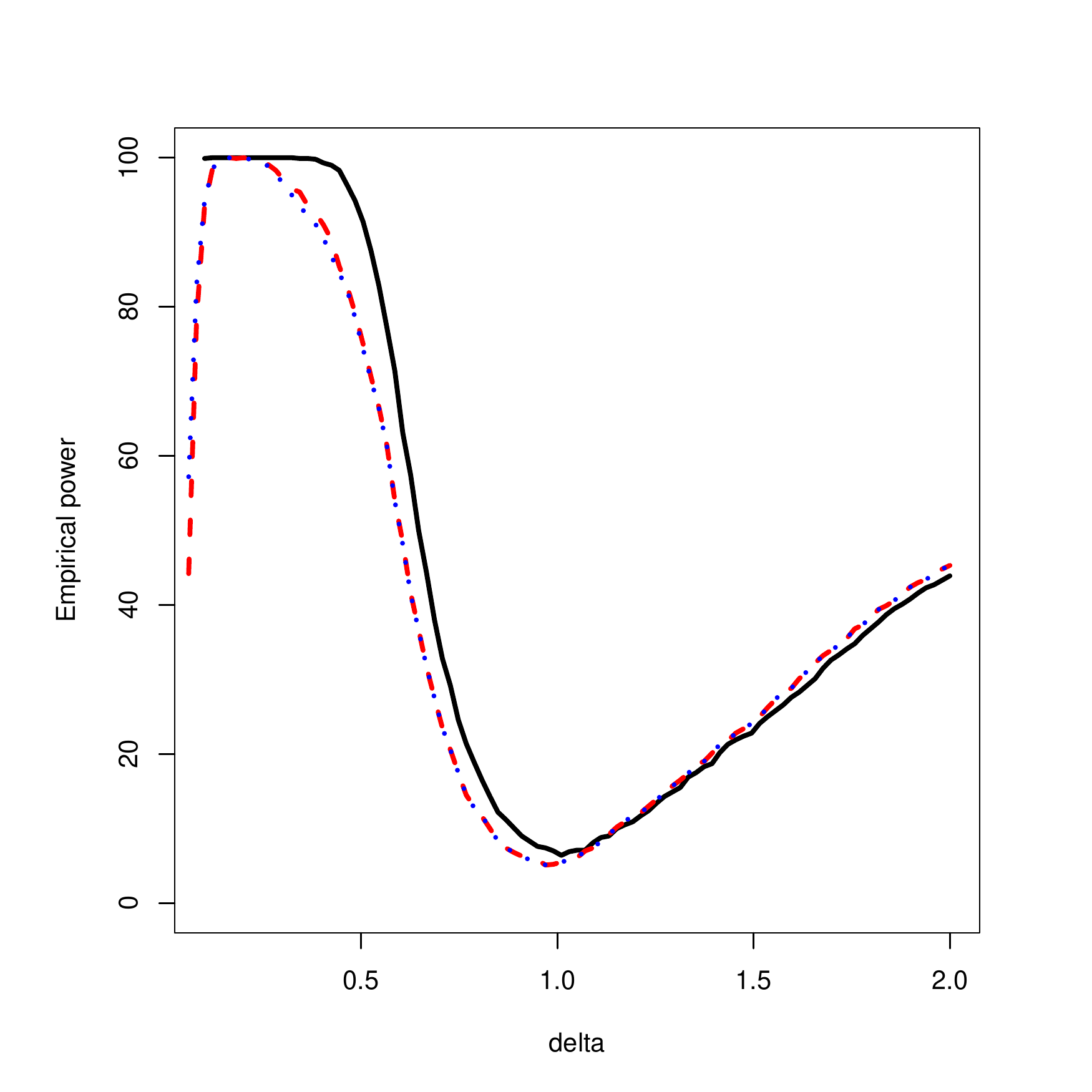}
  }
  \subfigure[$n = 100$]{
    \includegraphics[width = 0.26\linewidth]{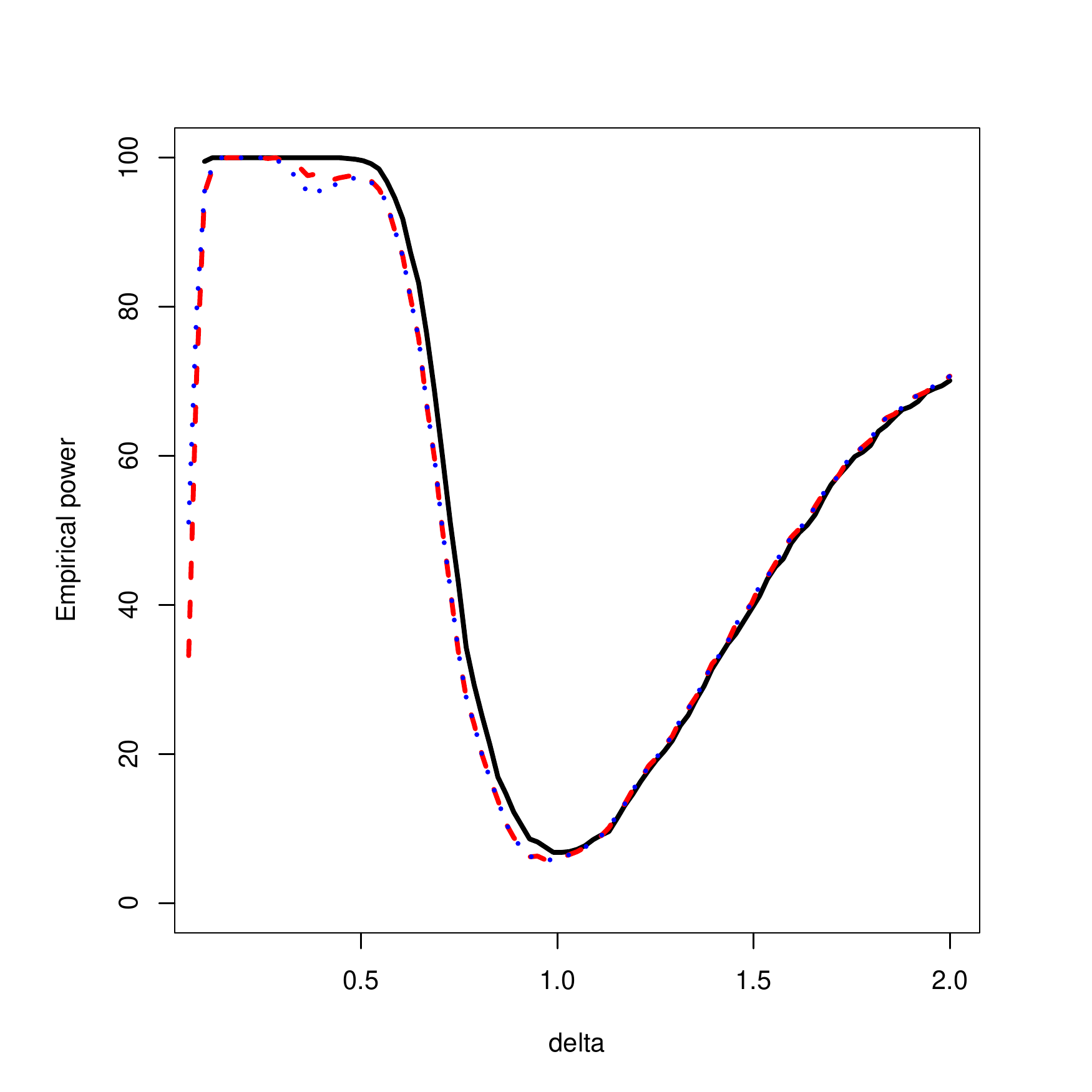}
  }
  \subfigure[$n = 20$]{
    \includegraphics[width = 0.26\linewidth]{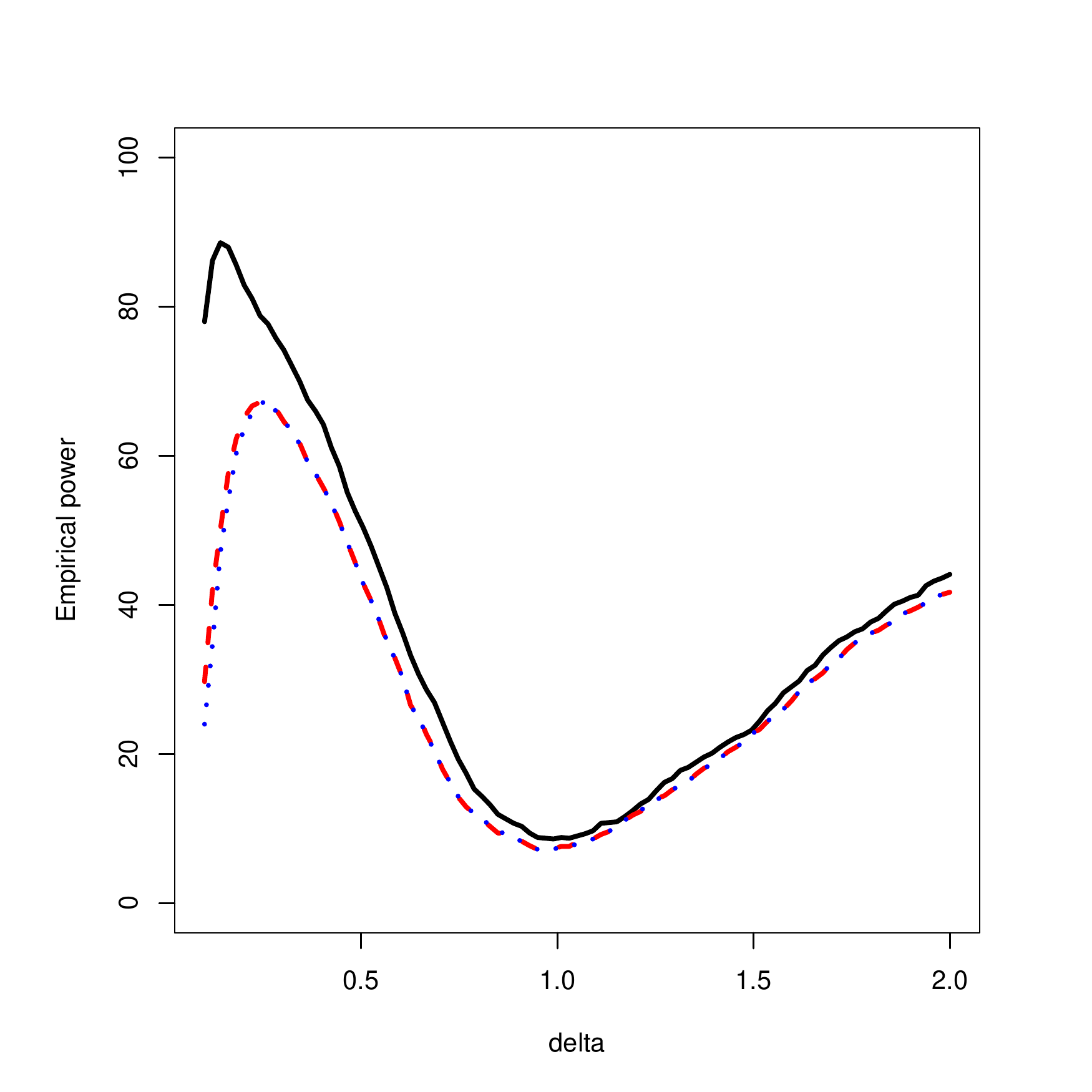}
  }
  \subfigure[$n = 50$]{
    \includegraphics[width = 0.26\linewidth]{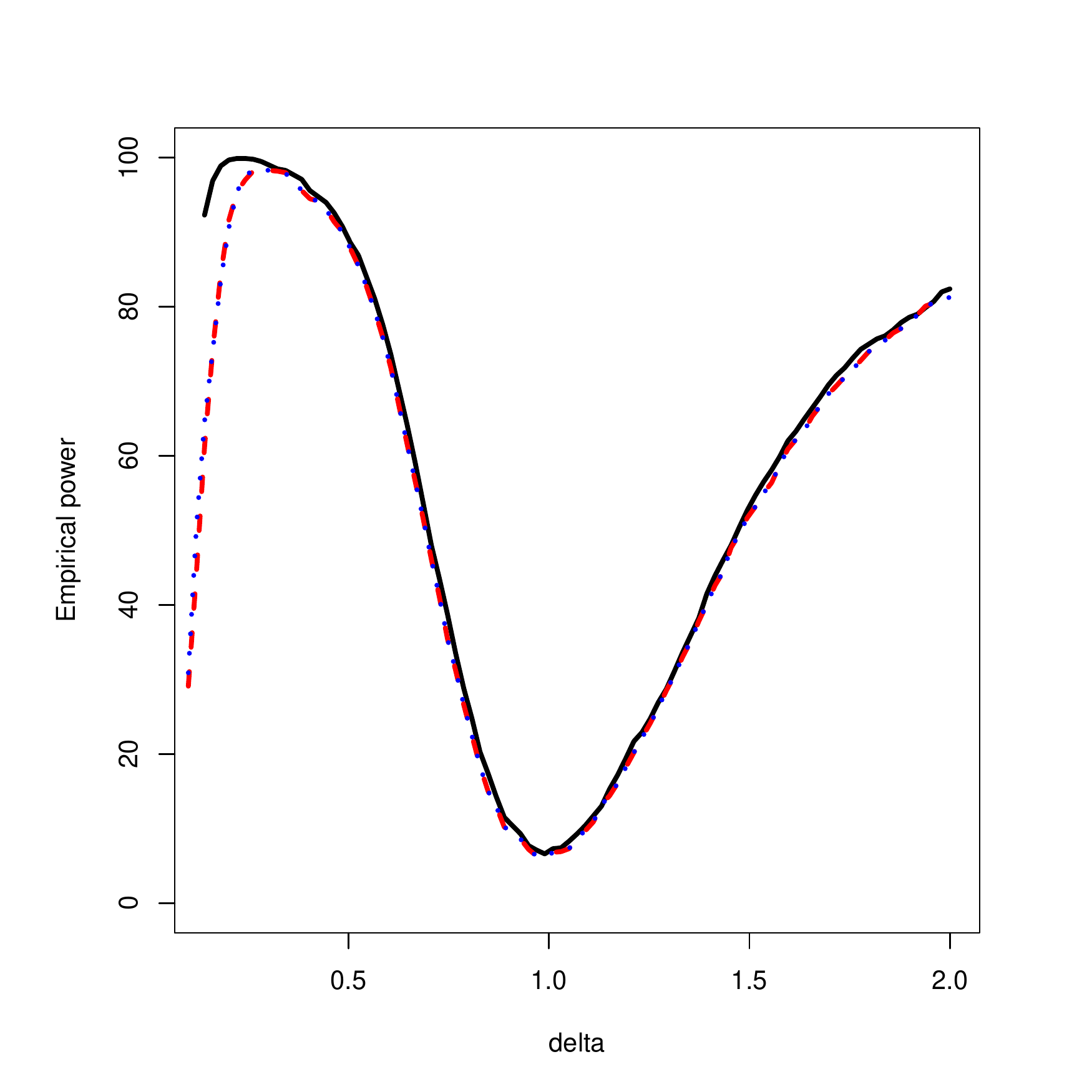}
  }
  \subfigure[$n = 100$]{
    \includegraphics[width = 0.26\linewidth]{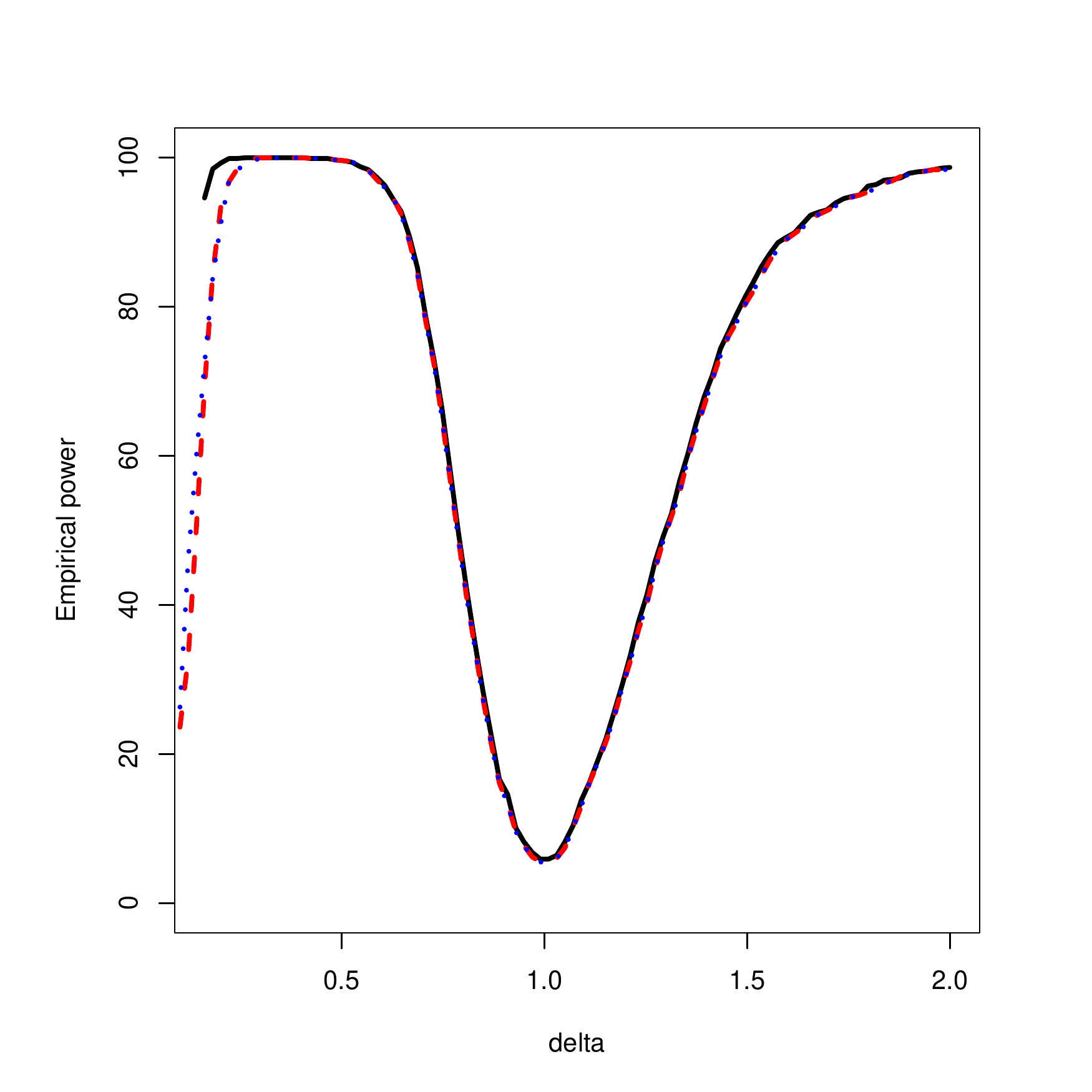}
  }
  \caption{Empirical power of $LM$ (solid line), $BF^A$ (dashed line) and $BF^B$ (dotted line)
  based in 1000 Monte Carlo replications. Panels are organized by each Scenario as described
  in Table \ref{tab:exp1}.}\label{fig:exp2}
\end{figure*}

\subsection{Setup}
We consider the model specification
\[
  \bm{y} = \bm{1}_n\beta_1 + \bm{x}_2\beta_2 + \exp(\bm{x}_3\beta_3) + \bm{\varepsilon},
\]
where $\bm{1}_n$ is a $n$-vector of ones, $\bm{x}_j\sim \mathsf{N}_n(\bm{0}, 0.16\,\bm{I})$,
$j=2,3$ and $\bm{\varepsilon}\sim \mathsf{N}_n(\bm{0}, 0.16\,\bm{I})$. Moreover, we
consider the following combinations of parameters
\[
  (\beta_1,\beta_2,\beta_3) \in \{(1,10,0.1),(1,5,0.2),(1,2,0.5),(1,1,1)\},
\]
and sample sizes $n\in\{20,50,100,500\}$. We test two equivalent null hypotheses
\begin{equation}\label{eq:H0A}
  H_0^A: \beta_2-\frac{1}{\beta_3} = 0,
\end{equation}
and
\begin{equation}\label{eq:H0B}
  H_0^B: \beta_2\beta_3 - 1 = 0.
\end{equation}
The number of Monte Carlo replications is set to 5000. In addition, we compute the
empirical power under the alternative hypotheses $H_1^A: \beta_2 - \delta/\beta_3
= 0$, and $H_1^B: \beta_2\beta_3 - \delta = 0$ for different values of $\delta$.
The \textsf{R} code to perform the simulations described in this section and some additional results are available
at \textsf{github}.\footnotemark[2]\footnotetext[2]{URL: \url{https://github.com/faosorios/BF_EE}}

\subsection{Comments on the simulations}

The results in Table \ref{tab:exp1} suggest that the $BF$ test works well in finite
samples even when the sample size is as small as $n=20$. In most of the considered
cases the $BF$ test outperforms the distance statistic $D$ as well as the $LM$ test.
It is worth noticing that, unlike $W$, the $BF$ test is not very sensitive to the
specification of the null hypothesis. Empirical power of the $BF$ and $LM$ tests is
displayed in Figure \ref{fig:exp2}. Using the $BF$ test may cause some loss of power.
However, as expected, as the sample size increases the empirical power of the $LM$ and
$BF$ tests becomes indistinguishable.

\section{Concluding remarks}\label{sec:conclusion}

In this paper we introduced a set of bilinear form tests for EE that may be considered
as a generalization of Terrell's gradient statistics \citep{Terrell:2002}. The asymptotic
distribution of the proposed tests is chi-square with degrees of freedom equal to the
number of restrictions. A Monte Carlo experiment shows that the $BF$ test works well
in finite samples and that it generally outperforms its competitors. Furthermore,
while the $BF$ test is not generally invariant to the specification of the null, its
finite sample performance seems to be only marginally affected by such a property.
It is worth noticing that, despite the favorable finite sample properties, the $BF$
test requires the estimation of the parameters of interest both under the null and
under the alternative. This feature may make it less attractive, from a computational
point of view, when compared to some of its competitors, e.g. the $LM$ test, that
only require the estimation of the parameters under the alternative. Nonetheless,
this type of development offers yet another alternative for carrying out hypothesis
tests in such general contexts as quadratic inference functions \citep{Qu:2000},
generalized empirical likelihood \citep{Newey:2004}, and maximum L$q$-likelihood
estimation \citep{Ferrari:2010}. We must emphasize that a detailed study of the
properties of local power and invariance of the $BF$ test deserves further exploration
along the lines of, for instance, \cite{Dagenais:1991} and \cite{Lemonte:2016}.

\section*{Acknowledgements}

The authors acknowledge the suggestions from an anonymous referee which helped to
improve the manuscript.

\appendix
\section{Proof of Proposition \ref{result:main}}\label{app:proof}

Following Property 24.16 in \cite{Gourieroux:1995}, we know that
\begin{align}
  \sqrt{n}(\what{\bm{\beta}}_n - \bm{\beta}_0) \stackrel{\sf D}{\to} \mathsf{N}_p(\bm{0},
  \bm{A}^{-1}\bm{B}\bm{A}^{-1}).
\end{align}
Then, by the delta method, we find that under $H_0:\bm{g}(\bm{\beta}_0) = \bm{0}$
\begin{equation}\label{eq:asymp-g}
  \sqrt{n}\,\bm{g}(\what{\bm{\beta}}_n) \stackrel{\sf D}{\to} \mathsf{N}_q(\bm{0},\bm{\Omega}).
\end{equation}
From Property 24.10 in \cite{Gourieroux:1995}, we have that $\widetilde{\bm{\beta}}_n$
and $\widetilde{\bm{\lambda}}_n$ are the solutions of the first order conditions of
the Lagrangian problem in Equation (\ref{eq:objective}):
\begin{align}
  \frac{\partial{Q}_n(\widetilde{\bm{\beta}}_n)}{\partial\bm{\beta}} - \bm{G}^\top(\widetilde{\bm{\beta}}_n)
  \widetilde{\bm{\lambda}}_n & = \bm{0} \label{eq:FO-1} \\
  \bm{g}(\widetilde{\bm{\beta}}_n) & = \bm{0} \label{eq:FO-2}
\end{align}
and $\widetilde{\bm{\beta}}_n$ is consistent. A Taylor expansion argument applied to
$\partial{Q}_n(\what{\bm{\beta}}_n)/\partial\bm{\beta}$ and $\partial{Q}_n(\widetilde{\bm{\beta}}_n)/\partial\bm{\beta}$
around $\bm{\beta}_0$, $\bm{A}_n(\bm{\beta}_0)$ $\stackrel{\sf a.s.}{\to}\bm{A}$
uniformly and simple calculations yield
\begin{equation}\label{eq:equiv-g}
  \sqrt{n}\,\bm{g}(\what{\bm{\beta}}_n) = \bm{G}\{-\bm{A}\}^{-1}\sqrt{n}\,\frac{\partial{Q}_n(\widetilde{\bm{\beta}}_n)}
  {\partial\bm{\beta}} + o_{\sf a.s.}(1).
\end{equation}
From the first order condition (\ref{eq:FO-1}),
\begin{equation}\label{eq:equiv-score}
  \sqrt{n}\,\frac{\partial{Q}_n(\widetilde{\bm{\beta}}_n)}{\partial\bm{\beta}}
  = \bm{G}^\top(\widetilde{\bm{\beta}}_n)\sqrt{n}\,\widetilde{\bm{\lambda}}_n,
\end{equation}
we obtain that
\[
  \sqrt{n}\,\widetilde{\bm{\lambda}}_n = [\bm{G}  \{-\bm{A}\}^{-1}\bm{G}^\top]^{-1}
  \sqrt{n}\,\bm{g} (\what{\bm{\beta}}_n) + o_{\sf a.s.}(1).
\]
Then, using (\ref{eq:asymp-g}), we find
\begin{equation}\label{eq:asymp-lambda}
  \sqrt{n}\,\widetilde{\bm{\lambda}}_n \stackrel{\sf D}{\to} \mathsf{N}_q(\bm{0},
  \bm{S}^{-1}\bm{\Omega}\bm{S}^{-1}).
\end{equation}
Let $\bm{\Omega} = \bm{R}\bm{R}^\top$ where $\bm{R}$ is a nonsingular $q\times q$ matrix.
Then, using standardized versions of (\ref{eq:asymp-g}) and (\ref{eq:asymp-lambda}), it
follows that
\begin{align*}
  BF_1 & = \{\bm{R}^{-1}\bm{S}\sqrt{n}\,\widetilde{\bm{\lambda}}_n\}^\top
  \bm{R}^{-1}\sqrt{n}\,\bm{g}(\what{\bm{\beta}}_n) \\
  & = n\widetilde{\bm{\lambda}}{}_n^\top\bm{S}\bm{\Omega}^{-1}\bm{g}(\what{\bm{\beta}}_n)
  \stackrel{\sf D}{\to} \chi^2_q.
\end{align*}
The proof for $BF_2$ and $BF_3$ follows from the equivalences
\[
  \sqrt{n}\,\bm{g}(\what{\bm{\beta}}_n) =\bm{G} \sqrt{n}(\what{\bm{\beta}}_n
  - \widetilde{\bm{\beta}}_n) + o_{\sf a.s.}(1)
\]
and
\[
  \sqrt{n}\,\widetilde{\bm{\lambda}}_n = \sqrt{n}\{\bm{G}^+\}^\top
  \frac{\partial{Q}_n(\widetilde{\bm{\beta}}_n)}{\partial\bm{\beta}}.
\]
The proof for $BF_4$, $BF_5$ and $BF_6$ follows by additionally assuming $\bm{B} =
-\bm{A}$. Finally, the proof for $BF_7$ uses the fact that $\bm{P}_G\bm{\Omega}\bm{P}_G
= \bm{\Omega}$.

\section{Proof of Proposition \ref{result:equiv}}\label{app:equiv}

From (\ref{eq:equiv-g}) and (\ref{eq:equiv-score}), we have that
\begin{align*}
  \sqrt{n}\,\bm{g}(\what{\bm{\beta}}_n) & = \bm{G}
  \{-\bm{A}\}^{-1}\bm{G}^\top\,\sqrt{n}\,\widetilde{\bm{\lambda}}_n+o_{\sf a.s.}(1) \\
  & = \bm{S}\,\sqrt{n}\,\widetilde{\bm{\lambda}}_n + o_{\sf a.s.}(1),
\end{align*}
and this implies that,
\begin{align*}
  BF & = \sqrt{n}\,\widetilde{\bm{\lambda}}{}_n^\top\bm{S}\bm{\Omega}^{-1}\,\sqrt{n}\,
  \bm{g}(\what{\bm{\beta}}_n) \\
  & = \sqrt{n}\,\widetilde{\bm{\lambda}}{}_n^\top\bm{S} \bm{\Omega}^{-1}\bm{S}\,\sqrt{n}\,
  \widetilde{\bm{\lambda}}_n + o_{\sf a.s.}(1).
\end{align*}
By using the asymptotic distribution given in Equation (\ref{eq:asymp-lambda}), the
proposition is verified.



\section*{References}

\begin{thebibliography}{10}
  \bibitem[Boos(1992)]{Boos:1992}
    Boos, D.D., 1992.
    On generalized score tests.
    The American Statistician 46, 327-333.

  \bibitem[Dagenais and Dufour(1991)]{Dagenais:1991}
    Dagenais, M.G., Dufour, J.-M., 1991.
    Invariance, nonlinear models, and asymptotic tests.
    Econometrica 59, 1601-1615.

  \bibitem[Ferrari and Yang(2010)]{Ferrari:2010}
    Ferrari, D., Yang, Y., 2010.
    Maximum L$q$-likelihood estimation.
    The Annals of Statistics 38, 753-783.

  \bibitem[Gourieroux and Monfort(1995)]{Gourieroux:1995}
    Gourieroux, C., Monfort, A., 1995.
    Statistics and Econometrics Models: Testing, Confidence Regions, Model Selection
    and Asymptotic Theory. Vol. 2.
    Cambridge University Press.

  \bibitem[Gregory and Veall(1985)]{Gregory:1985}
    Gregory, A.W., Veall, M.R., 1985.
    Formulating Wald tests of nonlinear restrictions.
    Econometrica 53, 1465-1468.

  \bibitem[Hansen(2006)]{Hansen:2006}
    Hansen, B.E., 2006.
    Edgeworth expansions for the Wald and GMM statistics for nonlinear restrictions.
    In: Corbae, D., Durlauf, S.N., Hansen, B.E. (Eds.), Econometric Theory and Practice:
    Frontiers of Analysis and Applied Research. Cambridge, 9-35.

  \bibitem[Hayashi(2000)]{Hayashi:2000}
    Hayashi, F., 2000.
    Econometrics.
    Princeton University Press.

  \bibitem[Lafontaine and White(1986)]{Lafontaine:1986}
    Lafontaine, F., White, K.J., 1986.
    Obtaining any Wald statistic you want.
    Economics Letters 21, 35-40.

  \bibitem[Lemonte(2016)]{Lemonte:2016}
    Lemonte, A., 2016.
    The Gradient Test: Another Likelihood-based Test.
    Academic Press, Amsterdam.

  \bibitem[Magnus and Neudecker(2007)]{Magnus:2007}
    Magnus, J.R., Neudecker, H., 2007.
    Matrix Differential Calculus with Applications in Statistics and Econometrics.
    Wiley, New York.


  \bibitem[Newey and Smith(2004)]{Newey:2004}
    Newey, W.K., Smith, R.J., 2004.
    Higher order properties of GMM and generalized empirical likelihood estimators.
    Econometrica 72, 219-255.

  \bibitem[Qu et al.(2000)]{Qu:2000}
    Qu, A., Lindsay, B.G., Li, B., 2000.
    Improving generalized estimating equations using quadratic inference functions.
    Biometrika 87, 823-836.

  \bibitem[Rao(1948)]{Rao:1948}
    Rao, C.R., 1948.
    Large sample tests of statistical hypotheses concerning several parameters with applications to problems of estimation.
    Proceedings of the Cambridge Philosophical Society 44, 50-57.

  \bibitem[Ruud(2000)]{Ruud:2000}
    Ruud, P.A., 2000.
    An Introduction to Classical Econometric Theory.
    Oxford University Press, New York.

  \bibitem[Terrell(2002)]{Terrell:2002}
    Terrell, G.R., 2002.
    The gradient statistic.
    Computing Science and Statistics 34, 206-215.

  \bibitem[Wei(1998)]{Wei:1998}
    Wei, B.-C., 1998.
    Exponential Family Nonlinear Models.
    Springer, Singapore.
\end{thebibliography}
\end{document}